\documentclass[12pt]{article}

\usepackage{graphicx,xspace,colortbl}

\usepackage{amsmath,amsthm,amsfonts}
\usepackage{fancybox}

\numberwithin{equation}{section}

\newtheorem{theorem}{Theorem}

\newtheorem{prop}[theorem]{Proposition}

\theoremstyle{definition}
\newtheorem{definition}{Definition}

\newtheorem{assumptions}{Assumptions}

\newtheorem{remark}[definition]{Remark}

\newtheorem{remarks}[definition]{Remarks}

\def\CD{{\cal D}}

\def\CF{{\cal F}}
\def\CG{{\cal G}}

\def\CW{{\cal W}}

\def\qed{\hfill$\sqcap\kern-8.0pt\hbox{$\sqcup$}$\\}
\def\beq{\begin{eqnarray}}
\def\eeq{\end{eqnarray}}
\def\beqq{\begin{eqnarray*}}
\def\eeqq{\end{eqnarray*}}
\def\beeq{\begin{eqnarray*}}
\def\eeeq{\end{eqnarray*}}

\def\be{\begin{equation}}
\def\ee{\end{equation}}

\newcommand{{\X}}{{\bf X}}
\newcommand{{\x}}{{\bf x}}
\newcommand{{\Z}}{{\bf Z}}
\newcommand{{\z}}{{\bf z}}
\newcommand{{\Y}}{{\bf Y}}
\newcommand{{\y}}{{\bf y}}
\newcommand{{\F}}{{\bf F}}

\title{Credit risk modeling
 using time-changed Brownian motion}
\author{T. R. Hurd\thanks{{Research supported by the
Natural Sciences and Engineering Research Council of Canada and
MITACS, Mathematics of Information Technology and Complex Systems
Canada}} \\ Dept. of Mathematics and Statistics\\ McMaster
University\\Hamilton ON L8S 4K1\\Canada}
\date{\today}

\begin{document}
\maketitle

\begin{abstract}

Motivated by the interplay between structural and reduced form credit models, we propose to model the firm value process as a time-changed Brownian motion that may include jumps and stochastic volatility effects, and to study the first passage problem for such processes. We are lead to consider modifying the standard first passage problem for stochastic processes to capitalize on this time change structure and find that the distribution functions of such ``first passage times of the second kind'' are efficiently computable in a wide range of useful examples. Thus this new notion of first passage can be used to define the time of default in generalized structural credit models. Formulas for defaultable bonds and credit default swaps are given that are both efficiently computable and lead to realistic spread curves.  Finally, we show that by treating joint firm value processes as dependent time changes of independent Brownian motions, one can obtain multifirm credit models with rich and plausible dynamics and enjoying the possibility of efficient valuation of portfolio credit derivatives. 
 \end{abstract}


\newpage

\section{Introduction \label{introduction}}

The structural approach to credit modeling, beginning with the works of Merton \cite{Merton74} and Black and Cox \cite{BlacCox76}, treats debt and equity as contingent claims (analogous to barrier options) on the firm's asset value process. This unification of debt with equity is conceptually satisfying, but in practice the approach sometimes leads to inconsistencies with intuition and observation, such as the zero short-spread property (a consequence of the predictable  nature of the default time), and, in Merton type models,  time inconsistency. The Black-Cox framework, while time consistent, also leads to technical difficulties when pushed to provide realistic correlations between different firms' defaults and with other market observables.  Both approaches tend to be too rigid to allow good fit to market data as well as effective hedging strategies.  

Reduced-form (or ``intensity-based'') modeling, introduced by Jarrow and Turnbull \cite{JarrTurn95}, has been successful in  providing remedies for some of these problematic aspects. It treats default as locally unpredictable, with an instantaneous hazard rate, but does away with the connection between default and the firm's asset value process. 

Subsequent developments have bridged the gap between reduced form and structural models. For example, the  model of Jarrow, Lando and Turnbull \cite{JarLanTur97} and its extensions \cite{Lando98,ArvGreLau99,HurdKuzn07} posit a continuous time Markov chain to replace the firm value process as a determinant of credit quality, while retaining the concept of hazard rate in the form of dynamically varying Markov transition rates. The time of default is the first-hitting time of the default state, an absorbing state of the Markov chain. The incomplete information approach of Jarrow and Protter \cite{JarrProt04} views reduced form methods as arising from structural models when the market has less than perfect information about the firm value. So-called hybrid models \cite{MadaUnal00} (see also \cite{CarrWu05}) seek to tighten the connection with structural models by allowing the hazard rate to depend on the firm's equity value (stock price), and allow the stock price to jump to zero at the time of default. 

Another useful way to generalize the structural framework is to consider other classes of stochastic processes for the firm value. This line of work can potentially parallel a great body of work that extends stock price return models by allowing jumps and stochastic volatility. While this can be quite successful in the Merton framework, the Black-Cox framework faces the technical difficulty associated with the first passage problem. Black-Cox style structural credit models  using  jump-diffusion processes  to model the firm value have been studied in  \cite{Zhou01b,RufSche06}. Because of the difficulty in solving the first passage problem, their models are hard to compute, but these studies do demonstrate the viability of the approach in curing some of the deficiencies of the classic Black-Cox model by adding flexibility and the possibility of unpredictable defaults.  As for exact formulas for first passage, results based on fluctuation theory \cite{Bingham75} and Wiener-Hopf factorization \cite{Bertoin96} are known. Kou and Wang \cite{KouWang03} solve the first passage problem for a specific class of jump-diffusion process, and Chen and Kou \cite{ChenKou05} use those results to extend the Black-Cox firm value model and the Leland-Toft model  \cite{LelaToft96} for the optimal capital structure of the firm. 

The purpose of the present paper is to explore time-changed Brownian motions (TCBM) for their potential to be used for consistent modeling of a firm's asset value process and the firm's time of default. We aim to retain flexibility (to be able to match a wide range of possible credit spread curves), computational tractability (to permit efficient option valuation), and logical consistency with the paper of Black and Cox (by treating default as a first passage time for the firm value to hit a default threshold).

Many authors have used TCBMs as models of log stock returns (a notable review paper is \cite{GemMadYor01}), and their great flexibility is by now well known. When the time change is an independent L\'evy process (L\'evy subordinator), one obtains well known models such as the variance gamma (VG) model and the normal inverse Gaussian (NIG) model.  Barndorff-Nielsen and Shephard \cite{BarnNielShep01} have introduced time change models where the time change is an integrated mean-reverting jump process, while important stochastic volatility models such as Heston's model \cite{Hest93} arise from time changes that are integrated mean-reverting diffusions.  The paper of \cite{CarGemMadYor03} surveys 25 different realizations of time changed L\' evy processes and shows how they perform in a calibration exercise to observed option pricing data. 

This paper focusses on the remaining difficulty in successfully implementing TCBMs to model the firm value in a Black-Cox framework, namely the efficient computation of first passage probabilities. Because of the difficulties that arise in computing the associated first passage distribution, and in analogy to the time-changed Markov chain models where the default state is an absorbing state, we are lead to propose a specific variation of first passage time applicable to time-changed Brownian motions, but not to general jump diffusions. This variation, which we call the {\it first passage time of the second kind}, is designed to be decomposable by iterated conditional expectation, and thus can be computed efficiently in cases of interest. This concept is not new, having been used for example by Moosbrucker \cite{Moosbr06} and Baxter \cite{Baxter06a} in their computations of basket credit derivatives, but to our knowledge its modeling implications have not yet been fully explored. 

Our purpose here is threefold. First we explore the mathematical structure of first passage times for time-changed Brownian motion, and provide a set of natural solvable examples that can be used in finance. By comparison of these examples with a range of existing stock price models, we thereby demonstrate the broad applicability of our framework to equity and credit modeling. Our second aim is to focus on structural models of credit where the firm value process is a general time-changed Brownian motion and the time of default is a first passage time of the second kind. We prove pricing formulas for defaultable zero coupon bonds  and credit default swaps, with and without stochastic recovery. This discussion demonstrates that time-changed Brownian motion can be the basis of single firm credit models consistent with the principles of no arbitrage, and with tractable valuation formulas for important derivative securities. Finally, we demonstrate how the single firm model can be extended to the joint default dynamics of many firms. Under a restrictive assumption on the correlation structure, analogous to the one-factor default correlation structure in copula models, we demonstrate the efficiency of valuation formulas for portfolio credit derivatives.  

To avoid obscuring our most important results by focussing on a too-specific application, we ask the reader's indulgence to postpone statistical work on the modeling framework to subsequent papers. While we are hopeful that statistical verification of the modeling assumptions on asset price datasets will  ultimately show the viability of our framework, such a verification must proceed one application at a time, and would take us too far in the present paper. 

In outline, the paper proceeds as follows. Section 2 introduces the probabilistic setting and the definition and basic properties of TCBMs. The first passage problem for TCBMs is addressed in Section 3. Since the standard first passage problem for TCBMs exhibits no simplification over first passage problem for general jump-diffusions, we introduce an alternative notion, called the first passage time of the second kind, that capitalizes on the time change structure. It is this notion that is used in all subsequent developments. Section 4  introduces the main categories of time changes, namely the L\'evy subordinators and the integrated mean-reverting jump-diffusions. These two families are in a sense complementary, and together provide a rich and tractable family of TCBMs that can be used as building blocks for more general models. Section 5 introduces the simplest structural credit models based on TCBMs, and runs through the valuation of some basic credit derivatives. Section 6 provides a brief numerical exploration of the single firm model. The multifirm extension is addressed in Section 7. We find that computational tractability strongly suggests that while the time change processes for different firms may (indeed should) be correlated, the underlying Brownian motions must be taken independent firm by firm.

\section{Time-changed Brownian motion}

Let $(\Omega, \CF, \mathbb{F}, P)$ be a filtered probability space that supports a Brownian motion $W$ and an independent strictly increasing c\'adl\'ag process $G$ with $G_0=0$, called the {\it time-change}. $P$ may be thought of as either the physical or risk-neutral measure. Let $X_t=x+\sigma W_t+\beta\sigma^2 t$ be the Brownian motion starting at $x$ having constant drift $\beta\sigma^2$ and volatility $\sigma>0$.  

\begin{definition} The  {\it time-changed Brownian motion (TCBM)}  generated by $X$ and $G$ is defined to be the process 
\be L_t=X_{G_t}, \quad t\ge 0.
\ee
To eliminate a redundant parameter, we normalize the speed of the time change:
\[ \lim_{T\to\infty}T^{-1} E[G_T ]=1.\]
\end{definition}
In addition to the ``natural'' filtration of the TCBM, $\CF_t=\sigma\{X_s, G_u: s\le G_t,u\le t\}$, one can also consider the subfiltration $\CG_t=\sigma\{G_s: s\le t\}$
and the Brownian filtration $\CW_t=\sigma\{W_s: s\le t\}$.

\begin{remarks}\begin{enumerate}
  \item 
Since the above definition requires $X$ and $G$ to be independent, it is somewhat more restrictive than notions of TCBM studied by others. However, as has been amply demonstrated by finance researchers such as \cite{GemMadYor01,CarGemMadYor03}, this family of processes offers a promising degree of versatility and tractibility when used to model equities. We will see in what follows how this family can be useful in credit risk modeling. 
 \item It is a point of philosophy to think of the time change as a reflection of the impact of the market on individual firms, in the same way that in reduced form models, the default intensity $\lambda$ lies in the market filtration. We think of the processes $X$ as firm specific generalizations of the default indicator function, while the time change generalizes the default intensity. 
\end{enumerate}\end{remarks}
The {\it characteristic functions} $\Phi$ for any $0\le s\le t$ and $u\in\CD$, $\CD$ a domain in $\mathbb{C}$ are defined to be
\beq
\Phi^X_s(u,t) &=& E[e^{iu(X_t-X_s)}|\CW_s]= e^{i\sigma^2(\beta u+iu^2/2)(t-s)} ,\nonumber\\
\Phi^G_s(u,t) &=& E[e^{iu(G_t-G_s)}|\CG_s], \nonumber\\
\Phi_s^L(u,t) &=& E[e^{iu(L_t-L_s)}|\CF_s].
\eeq
and real variable versions of these functions, in the form of the {\it Laplace exponents} $\psi(u)=-\log\Phi(iu)$, are also of interest. All are to be understood as processes in the variable $s$.
A simple calculation gives an essential formula 
\beq \Phi_s^L(u,t) &=& E[E[e^{iu(X_{G_t}-X_{G_s}0}|\CF_s\vee\CG_t]|\CF_s]\nonumber
\\
&=&E[\Phi_{G_s}^X(u,G_t)|\CG_s] =\Phi_s^G(\sigma^2(\beta u+iu^2/2),t).
\label{CharFnL}
\eeq
This formula reflects a kind of ``doubly stochastic'' or  ``reduced form'' property of TCBMs, already familiar in credit risk modeling: general $\CF$ expectations involving default can be ``reduced'' to expectations in the   ``market filtration'' $\CG_s$.

Useful ``solvable models'' arise when $\Phi_s^G$ and hence $\Phi_s^L$ are explicit deterministic functions of an underlying set of Markovian variables. Explicit characteristic functions not only lead to formulas for moments $m^{(k)}=E[L_t^k]$ and cumulants $c^{(k)}$ for $k=1,2,\dots$ but are also useful for extracting more detailed properties of the process $L$. An important algebraic aspect of TCBM is their natural composition rules. If $G, H$ are two independent time changes then $G+H$, $G\times H$ and $G\circ H$ are also TCBMs, and one has results such as 
$\Phi_s^{G+H}=\Phi_s^G\times \Phi_s^H$ and $\Phi_s^{G\circ H}=E[\Phi_{H_{s}}^G(u,H_t)|\CF_{s}]$.

\section{First passage distributions}

The first passage time for a semimartingale $L_t$ starting at a point $L_0=x>0$ to hit zero is the stopping time 
  \be  t^{(1)}=\inf\{t|L_t\le 0\}.\ee
The first passage problem for general semimartingales is difficult to deal with, and few applications have been realized. When applied to the nice subclass of TCBMs, it turns out that first passage times do not respect the reduced form property, and even in this setting they remain difficult to compute. On the contrary, when $L_t$ is a TCBM process, the following different definition of first passage time, which we call the {\it first passage time of the second kind}, is both natural and compatible with the ``reduced form'' property. Consequently we shall see that it can be easily implemented in credit modeling. 
\begin{definition} The {\it first passage time of the second kind} of the TCBM $L_t=X_{G_t}$ is the stopping time
\be t^{(2)}=\inf\{t|G_t\ge t^*\},\ee
  where $t^*=\inf\{t|X_t\le 0\}$. 
\end{definition}  
\begin{remarks}
\begin{enumerate}
\item  As shown in \cite{HurdKuzn08}, $t^{(2)}$ can be viewed as an approximation of the usual first passage time $t^{(1)}$ with $ t^{(1)}\ge t^{(2)}$. When $G$ is a continuous process, the two definitions coincide. 
  \item When the time change is a pure jump process with unpredictable jumps, both stopping times are totally inaccessible. In general, they can be written as the minimum of a predictable stopping time and a totally inaccessible stopping time.

\end{enumerate}
\end{remarks}

An essential first step in constructing tools for studying $t^{(2)}$ is to collect  ``structure'' functions associated with first passage $t^*$ for drifting Brownian motion $X_t=x+\sigma W_t+\beta\sigma^2 t$. The following formulas are well known (see for example \cite{BoroSalm96}):
\begin{enumerate} 
\item The cumulative distribution function for the first passage time of drifting Brownian motion is 
\be \label{firstpassBM}
P(t,x,\sigma,\beta):=E_{x}[{\bf 1}_{\{t^{*}\le t\}}]=N\left(\frac{-x-\beta\sigma^2 t}{\sigma\sqrt{t}}\right)+e^{-2\beta x}N\left(\frac{-x+\beta\sigma^2 t}{\sigma\sqrt{t}}\right).
\ee
\item For any $u>-\beta^2\sigma^2/2$, the Laplace exponent of $t^{*}$  is
\beq
\psi(u, x,\sigma, \beta)&:=&-\log E_x[e^{-u t^{*}}{\bf 1}_{\{ t^{*}<\infty \}}]\nonumber\\
&=&-\log\left[\int^\infty_0 e^{-ut}\left(\frac{\partial P(t,x,\sigma,\beta)}{\partial t}\right) dt\right]=x(\beta+\sqrt{\beta^2+2u/\sigma^2}).
\label{Fourier}
\eeq
\item The joint distribution function $E_x[{\bf 1}_{\{ t^{*}> t\}}{\bf 1}_{\{ X_t\ge \ell\}}]$ is given for $\ell\ge 0$ by
\be \label{jointcdf}
N\left(\frac{x-\ell+\beta\sigma^2 t}{\sigma\sqrt{t}}\right)-e^{-2\beta x}N\left(\frac{-x-\ell+\beta\sigma^2 t}{\sigma\sqrt{t}}\right)
\ee
\end{enumerate}

These elegant formulas of Brownian motion are needed in the theory of the second kind of passage problem, for which the structure functions of $t^{(2)}$ are computable via an intermediate conditioning. Thus, for example, its cumulative distribution function (CDF) is given by:
\beq\label{p2formula} P^{(2)}(t, x)&:=&E_{x}[{\bf 1}_{\{t^{(2)}\le t\}}]=E[E_x[{\bf 1}_{\{t^*\le G_t\}}|\CG_\infty]]\\&=&\int^\infty_0 P(y, x,\sigma,\beta)\rho_t(y) dy\nonumber\eeq
where $\rho_t$ is the density of $G_t$ and the function  $P$ is given by \eqref{firstpassBM}. 

While this formula can sometimes be used directly, in the many cases of interest where the Laplace exponent $\psi(u,t)$ of the time change $G_t$ is given in closed form, $P^{(2)}$ and other structure functions can be given a more useful Fourier representation.
\begin{prop}\label{P2prop} For any $x>0$ let  $L_t=X_{G_t}, X_t=x+\sigma W_t+\beta\sigma^2t$ be a TCBM where $G_t$ has Laplace exponent $\psi(u,t):=-\log E[e^{-uG_t}]$.  Then
\begin{enumerate}
  \item For any $t>0$ and $\epsilon\in\mathbb{R}$ the function $E_x[{\bf 1}_{\{t<t^{(2)}\}}\delta(L_{t}-\ell)]$ is given by
  \be\label{lemmano1} {\bf 1}_{\{\ell>0\}}\frac{e^{\beta(\ell-x)}}{2\pi }\int_{\mathbb{R}+i\epsilon}\left[e^{iz(\ell-x)}-e^{iz(\ell+x)}\right]e^{-\psi(\sigma^2(z^2+\beta^2)/2,t)}dz
  \ee
  while $E_x[{\bf 1}_{\{t\ge t^{(2)}\}}\delta(L_{t}-\ell)] $ is given by
  \be
   \label{condpdf}\frac{e^{\beta(\ell-x)}}{2\pi }\int_{\mathbb{R}+i\epsilon}e^{iz(x+|\ell|)}e^{-\psi(\sigma^2(z^2+\beta^2)/2,t)}dz.
  \ee
  \item For any $t>0$ the CDF $P^{(2)}(t,x)$ is given by
  \be \label{P2}
e^{-2\beta x}{\bf 1}_{\beta>0}+{\bf 1}_{\beta\le 0}- \frac{e^{-\beta x}}{\pi}\int^\infty_{-\infty}\frac{z\sin(zx)}{z^2+\beta^2}e^{-\psi(\sigma^2(z^2+\beta^2)/2,t)} dz, \ee
while   the characteristic function $
E_x[{\bf 1}_{\{t<t^{(2)}\}}e^{-\beta L_{t}+ikL_t}]$ is given for any $k$ in the upper half plane by
\be\frac{e^{-\beta x}}{2\pi }\int_{\mathbb{R}}\left[\frac{i}{k-z}-\frac{i}{k+z}\right]e^{izx}e^{-\psi(\sigma^2(z^2+\beta^2)/2,t)}dz.
  \label{phi2formula}
  \ee
   
\end{enumerate}
\end{prop}

\begin{remark}
The formulas in this proposition are all explicit Fourier integrals in the variable $x$ involving the Laplace exponent. This is a key advantage over a formula like \eqref{p2formula} in at least two respects. Firstly, the Laplace exponent is in many cases given explicitly while the density $\rho$ is not. Secondly, compared to a generic numerical integration, the fast Fourier transform (FFT) algorithm is  computationally efficient and comes with powerful error estimates as described in \cite{Lee04}. 
\end{remark}

\bigskip\noindent{\bf Proof of Proposition \ref{P2prop}:\ }  To show \eqref{lemmano1}, we note that
\[ E_x[{\bf 1}_{\{t<t^{(2)}\}}\delta(L_{t}-\ell)] =E[E_x[{\bf 1}_{\{G_t<t^*\}}\delta(X_{G_t}-\ell)|\CG_t]]
\]
where the inner expectation can be evaluated by differentiating \eqref{jointcdf} and using a standard Gaussian integral that holds for any $\epsilon$ and $t>0$: 
\be\label{gaussint}\frac1{\sigma\sqrt{2\pi t}}e^{-x^2/(2\sigma^2t)}=\frac1{2\pi}\int_{\mathbb{R}+i\epsilon} e^{-izx-z^2\sigma^2t/2}dz.
\ee
This then leads to \beeq
E_x[{\bf 1}_{\{t<t^{(2)}\}}\delta(L_{t}-\ell)]&=&E\left[{\bf 1}_{\{\ell\ge 0\}}\frac{e^{\beta(\ell-x)-\beta^2\sigma^2G_t/2}}{2\pi}\int_{\mathbb{R}+i\epsilon} e^{iz\ell}\left[e^{-izx}-e^{izx}\right]e^{-z^2\sigma^2G_t/2}dz\right]
\\&=&
{\bf 1}_{\{\ell\ge 0\}}\frac{e^{\beta(\ell-x)}}{2\pi}\int_{\mathbb{R}+i\epsilon} e^{iz\ell}\left[e^{-izx}-e^{izx}\right]E[e^{-\sigma^2(z^2+\beta^2)G_t/2}]dz\\&=&
{\bf 1}_{\{\ell\ge 0\}}\frac{e^{\beta(\ell-x)}}{2\pi}\int_{\mathbb{R}+i\epsilon} e^{iz\ell}\left[e^{-izx}-e^{izx}\right]e^{-\psi(\sigma^2(z^2+\beta^2)/2,t)}dz
\eeeq
where in the second step we have used  the Fubini Theorem  to interchange the integral and expectation. 

To prove \eqref{condpdf} we use similar logic to note that \beeq
&&E_x[{\bf 1}_{\{t\ge t^{(2)}\}}\delta(L_{t}-\ell)]+E_x[{\bf 1}_{\{t<t^{(2)}\}}\delta(L_{t}-\ell)]\\&&\hspace{1in}=\ E_x[\delta(L_{t}-\ell)]=E\left[\frac1{2\pi}\int_{\mathbb{R}+i\epsilon} e^{-\beta(x-\ell)}e^{-iz(x-\ell)-z^2\sigma^2G_t/2-\beta^2 \sigma^2G_t/2}dz\right].
\eeeq

To prove \eqref{P2} we integrate \eqref{lemmano1} over $\ell\in\mathbb{R}^+$. To interchange the order of the $\ell$ and $z$ integrations we need to take $\epsilon>\beta$ if $\beta\ge 0$, but may take $\epsilon=0$ if $\beta<0$. Then we find 
\beeq
1-P^{(2)}(t,x)&=&\frac{e^{-\beta x}}{2\pi }\int_{\mathbb{R}+i\epsilon}\left[\int^\infty_0e^{\beta \ell +iz\ell}d\ell\right]\left[e^{-izx}-e^{izx}\right]e^{-\psi(\sigma^2(z^2+\beta^2)/2,t)}dz\\&=&
-\frac{e^{-\beta x}}{2\pi }\int_{\mathbb{R}+i\epsilon}\frac{1}{i(z-i\beta)}\left[e^{-izx}-e^{izx}\right]e^{-\psi(\sigma^2(z^2+\beta^2)/2,t)}dz  \eeeq
The integrand has a residue of $(2\pi i)^{-1}e^{-2\beta x} $ at $z=i\beta$, and when $\beta>0$ we need to apply the Residue theorem to deform the $z$-contour to the real axis. For all $\beta$, the resulting integrals over $\mathbb{R}$ can be manipulated into the required real form. The proof of \eqref{phi2formula} is straightforward under the condition on $k$. \qed

\section{Time change models}
TCBMs have been well studied as models of log-stock prices. We outline a general approach to building TCBMs, and provide a number of distinct types of time change that can be used as ``building blocks''. 
\subsection{L\'evy subordinated Brownian motions} \label{LSBMs}

These TCBMs arise by taking $G$ to be a L\'evy time change, that is, a {\it L\'evy subordinator}. Such TCBMs are examples of L\'evy processes, the general class of continuous time stochastic processes with stationary and independent increments.  Much of the analysis connected with a L\'evy process $L_t$  is based on its {\em characteristic triple} $(\tilde b,\tilde c, \rho)_h$, in terms of which its Laplace exponent takes the form 
\be \psi^{L}(u,t):=-\log E[e^{-uL_t}] = t\left[\tilde b u-{\tilde c}^2u^2/2+\int_{\mathbb{R}\setminus 0} [e^{-uy}-1-uyh(y)]\rho(y) dy\right].
\ee
Here $\rho$ is a measure on $\mathbb{R}\setminus 0$. For ease of exposition in what follows, we set the truncation function $h(y)$ to zero, which is permissible by adopting the restrictive condition that $|x|\wedge 1$ should be $\rho$-integrable. Our main results extend to the general case where $|x|^2\wedge 1$ is $\rho$-integrable. See  \cite{ContTank04} for general discussions of L\'evy processes.

The following result is Theorem 4.3 in \cite{ContTank04}, and  identifies the type of process that can be expressed as a  {\em L\'evy-subordinated Brownian motion} (LSBM) $L_t:=X_{G_t}$:
\begin{theorem} Supposing $L_0=x$, the following are equivalent statements:
\begin{enumerate}
  \item $L$ is a L\'evy process with characteristic triple $(\tilde b,\tilde c, \rho)_0$ where $\tilde c\ge 0$. The density $\rho$  is nowhere zero on $\mathbb{R}$ and for some $\beta$,  $e^{\beta\sqrt{z}}\rho(-\sqrt{z})=e^{-\beta\sqrt{z}}\rho(\sqrt{z})$ and is a completely monotone function. Furthermore $\tilde b=\beta\tilde c$. 
    \item  $L_t:=X_{G_t}$ for drifting Brownian motion $X_t=x+W_t+\beta t$ and $G$ a  L\'{e}vy subordinator with characteristic triple $(b, 0, \nu)_0, b\ge 0$, and $\nu$ a measure on $(0,\infty)$.
\end{enumerate}
\end{theorem}

Here are some examples time changes $G$ for $L_t=X_{G_t}, X_t=x+\sigma W_t+\beta\sigma^2 t$ that have been used in models of logarithmic stock returns: \begin{enumerate}
 \item The exponential model with parameters $(a,b,c)$ arises by taking $G$ to be the increasing process with drift $b> 0$  and jump measure $\nu(z)= ac e^{-az}, c, a>0$ on $(0,\infty)$. The Laplace exponent of $G_t$ is
 \[ \psi(u,t):=-\log E[e^{- u G_t}]=t[bu+uc/(a+u)].\]
 and the normalization condition is $b+c/a=1$.
  The resulting time-changed process $L_t:=X_{G_t}$ has triple  $(\beta b, b, \rho )_0$ with
 \[ \rho(y)=\frac{c}{\sqrt{\beta^2+2a}}e^{-(\sqrt{\beta^2+2a}-\beta)(y)^+-(\sqrt{\beta^2+2a}+\beta)(y)^-},\]
where  $(y)^+=\max(0,y), (y)^-=(-y)^+$. This forms a four dimensional subclass of the six-dimensional family of exponential jump diffusions applied to finance in  \cite{KouWang03}. 
  \item The VG model \cite{MadaSene90}  arises by taking $G$ to be a gamma process with drift defined by the characteristic triple $(b, 0, \nu)_0$ with $b\ge 0$ (often $b$ is taken to be $0$) and jump measure $\nu(z)=c e^{-az}/z, c, a>0$ on $(0,\infty)$. The Laplace exponent of $G_t$ is
 \[ \psi(u,t):=-\log E[e^{-uG_t}]=t[bu+c\log(1+u/a)].
  \]
  and the normalization condition is $b+c/a=1$.
  The resulting time-changed process has triple  $(\beta b, b, \rho )_0$ with
 \[ \rho(y)=\frac{c}{|y|}e^{-(\sqrt{\beta^2+2a}+\beta)(y)^+-(\sqrt{\beta^2+2a}-\beta)(y)^-}.\]
  \item The normal inverse Gaussian model (NIG) with parameters $\tilde\beta, \tilde \gamma$ \cite{Barn-Niel97} arises when $G_t$ is the first passage time for a Brownian motion with drift $\tilde\beta>0$ to exceed the level $\tilde \gamma t$. Then 
 \[ \psi(u,t)=\tilde\gamma t(\tilde\beta+\sqrt{\tilde\beta^2+2u})
 \]
 and  the normalization condition is $\tilde \gamma/\tilde \beta=1$.
  The resulting time-changed process has  Laplace exponent
 \[ \psi^L(u,t)=xu+t\tilde\gamma [\tilde \beta+\sqrt{\tilde\beta^2+2\tilde\beta u+u^{2}}].
 \]
\end{enumerate}

\subsection{Affine TCBMs}\label{ATCBM}

For our second important class of time changes, $G_t=\int^t_0\lambda_s ds$ has differentiable paths, and the corresponding TCBMs are diffusions (processes with continuous paths) which exhibit ``stochastic volatility''. We focus here on a class we call ATCBMs (``affine'' TCBMs), for which $\lambda$ is taken in the class of positive mean-reverting CIR-jump  processes introduced by \cite{DuffSing99}. We mention here two distinct examples:
\beq  
d\lambda^{(1)}_t &=& (a-b\lambda^{(1)})dt +\sqrt{2c\lambda_t^{(1)}} dW^{(1)}_t, a,b,c>0,\nonumber\\
d\lambda^{(2)}_t &=& -\tilde b \lambda^{(2)}dt + dJ_t.
\eeq
Here $J$ is taken identical to the exponential L\'evy subordinator with parameters $(\tilde a,0,\tilde c)$ defined in example 1 of the previous subsection. 

The essential computations for Laplace exponents $$\psi^{(i)}(u,t;\lambda):=-\log E[e^{-uG^{(i)}_t}|\lambda^{(i)}_0=\lambda], i=1,2$$ of such affine time changes are described in many papers. The following formulas are proved in the appendix of \cite{HurdKuzn07}:
\begin{prop}
The characteristic functions $\psi^{(i)}, i=1,2$, both have the affine form
  \be
\psi^{(i)}(u,t;\lambda)=A^{(i)}(u,t)+\lambda B^{(i)}(u,t).
  \ee
  The functions $A^{(i)}$ and $B^{(i)}$ are explicit:
 \begin{enumerate}
\item \beq\label{eq_CIR}
  \begin{cases}
 A^{(1)}(u,t)=-\kappa
_2+\Bigl(1+\frac{c}{\gamma}\kappa
_1\left(e^{\gamma t}-1\right)\Bigr)^{-1}\kappa
_2,
\\
B^{(1)}(u,t)=-a\kappa
_1t+\frac{a}{c} \log\left(1+\frac{c}{\gamma}\kappa
_1\left(e^{\gamma t}-1\right)\right),
 \end{cases}
  \eeq
  with constants $\kappa
_1,\kappa
_2$ and $\gamma$ given by
  \beq
 \begin{cases}
\gamma=\sqrt{b^2+4uc},\\
\kappa
_1=\frac{b+\gamma}{2c},\\
\kappa
_2=\frac{b-\gamma}{2c}.
\end{cases}
  \eeq
\item 
 \beq\label{eq_MR}
\begin{cases}
A^{(2)}(u,t)=\frac{u}{\tilde b}\left(1-e^{-\tilde bt}\right),\\
B^{(2)}(u,t)=\tilde ct-\frac{\tilde a\tilde c}{\tilde a\tilde b+u}\log\left(\frac{(\tilde a\tilde b+u)e^{\tilde bt}-u}{\tilde a\tilde b}
\right).
\end{cases}
  \eeq

 \end{enumerate}

\end{prop}

The ATCBM model with time change $\int^t_0\lambda_s^{(1)}ds$ is equivalent to the Heston stochastic volatility model for stock returns \cite{Hest93}, with zero correlation (hence zero leverage effect). Stock price models with time change $\int^t_0\lambda_s^{(2)}ds$, and extensions thereof, were introduced  in \cite{BarnNielShep01}.

\subsection{More general TCBMs}

Two different ways of combining time changes have been studied that preserve the desirable property that the resulting Laplace exponent is explicitly known. The first is to add time changes. For example, a model that includes time jumps and stochastic volatility with both a diffusive and jump component arises if we take 
\[ G_t=\int^t_0(\lambda_s^{(1)}+\lambda_s^{(2)})ds + G^{(3)}_t\]
where $G^{(3)}_t$ is a L\'evy subordinator. 

The paper \cite{CarGemMadYor03} explores 25 models of the form $L_{H_t}$ where $L$ is a L\'evy process and $H_t=\int^t_0 \lambda_s ds$ is an independent time change similar to those mentioned in this section. In most cases they discussed, $L$ is itself of the form $X_{G_t}$, and so $L_{H_t}$ is a TCBM with time change $G\circ H$. Amongst this class, the paper finds several models of the stock price that capture very well the implied equity volatility surface.

From the large literature on such TCBMs as models for equities, it is clear that the class of TCBMs with explicit Laplace exponent is rich enough to describe a wide range of asset classes in finance. We shall now see how such processes can be used to build structural models of credit risk.  

\section{Structural credit models}
\label{debtmodel}
The structural credit modeling paradigm of Black and Cox \cite{BlacCox76} assumes that default of a firm is triggered as the debt holders exercise a ``safety covenant'' when the value of the firm falls to a specified level. It makes sense therefore to assume that the time of default is the time of first passage of the firm value process $V_t$ below a  specified lower threshold function $K(t)$. 

In this section, we outline how a Black-Cox credit framework can be built under the assumption that the log-leverage ratio $L_t=\log(V_t/K(t))$ is a TCBM. To demonstrate the flexibility of the approach, we make use of all the building blocks introduced so far, leading to a large number of parameters. A more realistic implementation would likely begin with a much more restricted specification. In analogy to the multifactor reduced form modeling framework of \cite{DuffSing99} we choose the approach in which independent time changes are combined together by addition rather than composition. The alternative route via composition is deserving of separate study. 

\begin{assumptions}
 \begin{enumerate}
  \item There is a vector $Z_t=[\tilde r_t,\lambda^{(1)}_t,\lambda^{(2)}_t]$ of independent processes with $\lambda^{(i)}$ chosen as in subsection \ref{ATCBM}. $\tilde r$ is a CIR process with Laplace exponent 
  $\psi^{\tilde r}(u,t)$ given in the form \eqref{eq_CIR}.
  \item The log-leverage process $L_t=\log(V_t/K(t))=X_{G_t}, X_t=x+\sigma W_t+\beta\sigma^2 t$ is a TCBM where the time change  is a convex combination of the building blocks of the previous section: 
  \be G_t=\alpha_1G^{(1)}_t+\alpha_2 G^{(2)}_t+\alpha_3G^{(3)}_t
   \label{TCfactors}
  \ee
  with $ 0\le \alpha_1,\alpha_2,\alpha_3\le 1=\alpha_1+\alpha_2+\alpha_3.$
  Here $G^{(i)}_t=\int^t_0\lambda^{(i)}_s ds,i=1,2 $ are defined as in Section \ref{ATCBM} with Laplace exponents $\psi^{(i)}(u,t;\lambda^{(i)})$ while $G^{(3)}$ is a L\'evy subordinator with Laplace exponent $\psi^{(3)}(u,t)$. We also assume that $\beta<0,\sigma>0$.
 \item The time of default is $t^{(2)}$, the first passage time of the second kind. 
  \item The spot interest rate is $ r_t=\tilde r_t+m_1\lambda^{(1)}_t+m_2\lambda^{(2)}_t$
  for non-negative coefficients $m_1,m_2$.
  \item A constant recovery fraction $R<1$ under the recovery of treasury mechanism  is paid on defaultable bonds at the time of default. (This is for simplicity:  \cite{HurdKuzn07} shows that we can allow $R_t$ to be a general affine process. We can also compute under the recovery of par assumption with somewhat more complex integrations.)
\end{enumerate}
\end{assumptions}

As previously mentioned, Black-Cox models using jump diffusions  are usually based on the standard first passage time, leading to technical difficulties that can be solved in only a restricted class of processes. Our innovation is to consider instead the second kind of first passage time, and thereby capitalize on a type of ``reduced form'' for computations that applies to general TCBMs.

\subsection{Bond pricing} The following proposition gives formulas for default probabilities and default-free and defaultable zero coupon bond prices.

\begin{prop} Let the initial credit state of the firm be specified by initial values $L_0=x$ and $Z_0=[\tilde r_0,\lambda^{(1)}_0,\lambda^{(2)}_0]$. Recall that $\beta<0$.
\begin{enumerate}
\item The probability that default occurs before $t>0$ is given by
\be P[t^{(2)}\le t]=1-\frac{e^{-\beta x}}{\pi}\int^\infty_{-\infty}\frac{z\sin(zx)}{z^2+\beta^2}\exp\left[-\sum_{i=1}^3\psi^{(i)}(\alpha_i\sigma^2(z^2+\beta^2)/2,t)\right] dz.
\ee
\item The time $0$ price $ P_0(T)$ of the default-free zero coupon bond with maturity $T$ is
\be P_0(T)=\exp\left[-\psi^{\tilde r}(1, T;\tilde r_{0})-\sum_{i=1,2}\psi^{(i)}(m_i,T;\lambda^{(i)}_0)\right].
\ee
  \item The time $0$ price  of the defaultable zero coupon bond with maturity $T$, under  constant fractional recovery of treasury, is $\bar P^{RT}_0(T)=(1-R)\bar P^{0}_0(T)+RP_0(T)$, where $\bar P^{0}_0(T)$ denotes the price of the zero recovery defaultable zero coupon bond, given by
  \beq \label{zrzcbond}
 &&\hspace{.5in} \bar P^{0}_0(T)= \frac{e^{-\beta x-\psi^{\tilde r}(1,T;\tilde r_0)}}{\pi}\int^\infty_{-\infty}\frac{z\sin(zx)}{z^2+\beta^2}\\&&\hspace{-.25in} \times\exp\left[-\sum_{i=1,2}\psi^{(i)}(m_i+\alpha_i\sigma^2(z^2+\beta^2)/2,T;\lambda^{(i)}_0)-\psi^{(3)}(\alpha_3\sigma^2(z^2+\beta^2)/2,T)\right] dz.\nonumber
   \eeq
 \end{enumerate}
\end{prop}

\begin{proof} We prove only the formula for $P^{0}_0(T)$: the other formulas are similar, but easier. The risk-neutral pricing formula gives
\beeq \bar P^{0}_0(T)&=&E\left[e^{-\int^{T}_0 r_s ds}{\bf 1}_{\{T< t^{(2)}\}}\right] \nonumber\\
&=&E\left[e^{-\int^T_0 r_s ds}E[{\bf 1}_{\{G_T<t^*\}}|\CG]  \right] \nonumber\\
&=&\frac{e^{-\beta x}}{\pi}\int^\infty_{-\infty}\frac{z\sin(zx)}{z^2+\beta^2}E\left[e^{-\int^T_0 r_s ds}e^{-\sigma^2(z^2+\beta^2)G_T/2}  \right]\ dz\eeeq
The third equality comes by using \eqref{P2} in the case when $G_t$ is deterministic, and using the Fubini Theorem to interchange the order of integration. The final form for \eqref{zrzcbond} now follows by decomposing  $\int^T_0 r_s ds$ and $G_T$ into their independent components and performing the resulting one dimensional expectations.\end{proof}

The above pricing formulas are explicit functions of the initial values $L_0=x, Z_0$; as time develops, prices are deterministic functions of the processes $L_{t}$ and $Z_{t}$. We adopt the point of view that $Z$ contains information about the drivers of general credit markets, while $L$ reflects firm specific information. 

\subsection{Credit default swaps}
We next consider an idealized CDS with unit notional and maturity $T>0$. The premium leg, paid continuously at a constant rate $S$ until $t^{(2)}\wedge T$, has the time $0$ price
\be S\ V(T)=SE\left[\int^T_0 e^{-\int^t_0 r_s ds}{\bf 1}_{\{t\le t^{(2)}\}}\ dt\right]= S\int^T_0 \bar P^0_0(t) dt
\ee
where the second equality comes by comparison to \eqref{zrzcbond}.
The default leg pays the fractional loss of treasury value $(1-R)P_0(t^{(2)})$ at the time of default if $t^{(2)}\le T$ and has the time $0$ value $W(T)=\frac{1-R}{R}(P_0(T)-\bar P^0_0(T))$. The CDS spread $S(T)$ is defined to be the value of $S$ that makes $SV=W$.

\section{Numerical results}

The structural credit modeling
 framework of the previous sections is designed with flexibility and computability in mind. Rather than embark here on a lengthy statistical investigation of promising specifications and their calibration to market data, instead, in this section we strip out the complexity, and simply exhibit a set of parametrizations of the VG TCBM model that generate plausible credit spread curves, thereby demonstrating the computational efficiency. 

We consider the credit framework for a pure geometric Brownian motion model (Model A) and three parametrizations of the VG model of \S \ref{LSBMs} (Models B,C,D), with parameters shown in Table 1. All four models are specified so that $L_t$ has $L_0=1.5$, fixed annualized variance $\sigma^2+\beta^2\sigma^4(2/a+1)=0.09$ and mean log rate of return $-\sigma^2/2$ (i.e. $\beta=-0.5$). 

In Figure 1, we compare the thirty year zero recovery yield spread and default probability density for these models. We observe that the yield spreads equalize as maturity increases, but show the completely different short time behaviour expected from the presence of jumps.  
Figure 2 shows the  thirty year zero recovery yield spreads in Model B for four firms which differ in their initial  distance-to-default values $L_0=0.3, 0.6, 1.0, 2.0$. We see here that firms with small $L_0$ (high default risk) can have decreasing spread curves, while the reverse is true if $L_0$ is large.

\begin{figure}[b]
\centering\includegraphics[scale=.8]{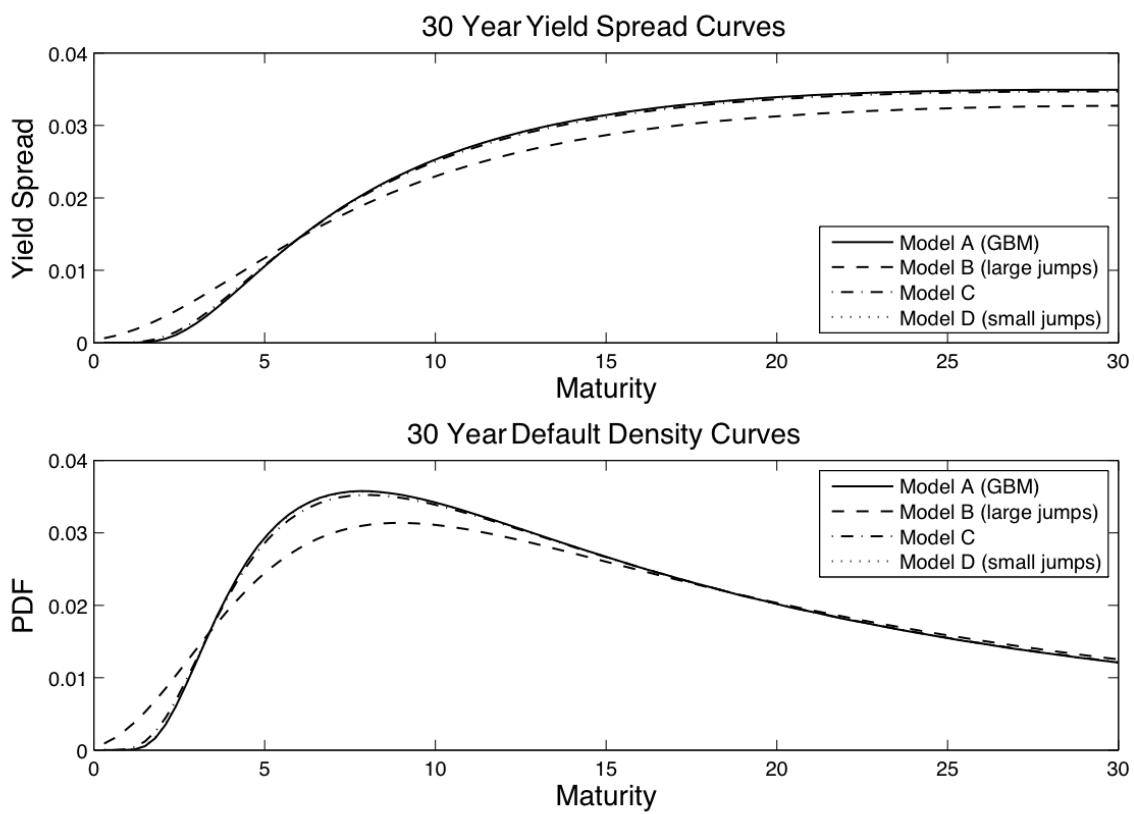}

\caption{Thirty year yield spread and default PDF curves for geometric Brownian motion model and three versions of the VG TCBM credit risk model.}
\end{figure}

\begin{figure}[b]
\centering\includegraphics[scale=.8]{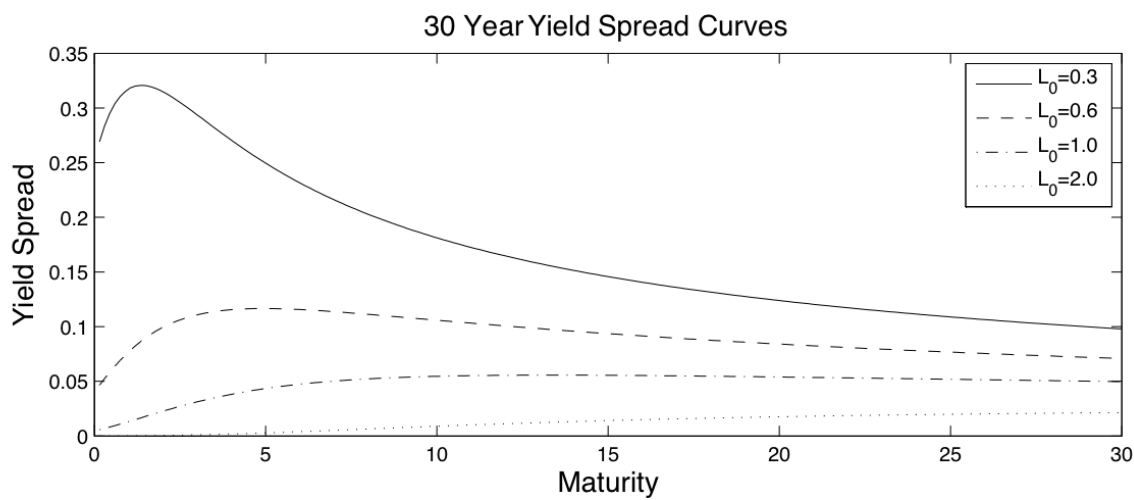}

\caption{Thirty year yield spread  for Model B with four different values $L_0=0.3, 0.6, 1.0, 2.0$. }
\end{figure}

\begin{table}[h]
\centering\begin{tabular}{|c||c|c|c|c|}
\hline 
&Model A&Model B&Model C&Model D 
\tabularnewline
\hline\hline 
$L_0$&1.5&1.5&1.5&1.5\tabularnewline\hline\hline 
$a$&1&1&10&100\tabularnewline
\hline
$b$&1&0&0&0\tabularnewline
\hline
$c$&0&1&10&100\tabularnewline
\hline 
$\beta$&-0.5&-0.5&-0.5&-0.5\tabularnewline
\hline
$\sigma^2$&0.09&0.0846&0.0877&0.0880\tabularnewline
\hline
\end{tabular}

\label{params}
\caption{Parameter values for the VG TCBM model.}
\end{table}

\section{A one factor multifirm structural model} 
A difficulty in structural credit risk modeling is finding a framework that extends naturally and efficiently to a large number of firms, while allowing for a rich default dependence structure. The present setup of time-changed Brownian motions is such a framework. Consider $M$ firms, where for each $m=1,2,\dots, M$, the $m$th firm is governed by its firm value process $V^m_t$, default trigger threshold $K^m(t)$ and log-leverage ratio process
\beq\label{ddk}
 L^m_t&=&\log V^m_t/K^m(t)=X^m_{G^m_t},\nonumber\\
 X^m_t&=&x_m+\sigma_mW^m_t+\beta_m\sigma_m^2t.
 \eeq
For the $m$th firm we take parameters $x_m, \beta_m,\sigma_m> 0$,  and a firm dependent time change  $G^m_t$. 

\begin{assumptions} The joint dynamics of multifirm defaults is determined by the first passage times $t_m^{(2)}
$ of the log-leverage ratio processes $L^m_t$.   The time change processes $G^m$ are given jointly in the ``one-factor'' form
\be\label{timechangek} G^m_t= \alpha_mG_t+(1-\alpha_m)H^m_t
\ee
with $\alpha_m\in[0,1]$ and time changes $G, H^m$ having the form given by \eqref{TCfactors}. Finally, we assume that  $G, H^1,\dots, H^M, X^1,\dots, X^M$ are mutually independent  processes.
\end{assumptions}

In this one-factor time change model, the maximal  correlation structure is obtained by setting each $\alpha_m=1$. However, since the underlying Brownian motions $X^m$ are independent, maximal correlation does not mean the defaults are fully correlated. 

The one factor model can be interpreted as a generalized Bernoulli mixing model, in the sense of \cite{BluOveWag03} and \cite{McNFreEmb05}, where the mixing random variable is $G_t$. That is, the default states of all firms at time $t$ are conditionally independent Bernoulli random variables, conditioned on $\CG_t:=\sigma\{G_s:s\le t\}$. If we define the conditional survival probability $F^m(x_m,G_t):=E_{x_m}[1_{\{G_t<t^{m*}\}}|\CG_t]$, then the following formula extends \eqref{P2} and is proved exactly the same way: 
\beq &&  \ \hspace{1.5cm}F^m(x,y)=\\
&&\hspace{-.5cm}\frac{e^{-\beta_m x}}{\pi}\int^\infty_{-\infty}\frac{z\sin(zx)}{z^2+\beta_m^2}\exp[{-\alpha_m\sigma_m^2(z^2+\beta_m^2)y/2-\psi^{H^m}((1-\alpha_m)\sigma_m^2(z^2+\beta_m^2)/2,y)}]dz. \nonumber \eeq
 
 Now, for any subset $\sigma\subset\{1,2,\dots, M\}$, the unconditional probability that the firms in default at time $t$ are precisely the firms in  $\sigma$ is given by
 \beq
 P[t_m^{(2)}
\le t, m\in\sigma; t_m^{(2)}
>t, m\notin\sigma]&=&\nonumber\\
&&\hspace{-1.5in}
 \int^\infty_0\prod_{m\in\sigma}\left(1-F^m(x^m,y)\right)\prod_{m\notin\sigma}F^m(x^m,y) \rho_t(y)dy,
 \eeq
 where $\rho_t$ is the distribution function of $G_t$.
 
There are by now well-known techniques that under the assumption of conditionally dependent defaults, reduce the computation of credit portfolio loss distributions and CDO tranches to intensive computation of the conditional survival probabilities $F^m(x,y)$.  

\section{Conclusions}

We have studied the first passage problem for a class of semimartingales that are important for financial modeling, namely Brownian motions time-changed by an independent time change process. It was seen that the first passage time of the second kind presents some key advantages over the standard definition of first passage time, particularly computational tractability and the possibility of extension to multi-dimensional processes. 

Based on these good properties, we defined a pure first passage structural model of default, and obtained computable formulas for the basic credit instruments, namely bonds and CDSs. The resultant formulas resolve a fundamental deficiency of the classic Black-Cox formula, namely the zero short spread property,  and provide needed flexibility to match details of yield spreads.

Finally, we outlined an extension to many firms in which dependence stems from systemic components to the time change, while the underlying Brownian motions are independent and firm specific. The resulting multifirm framework has a conditional independence structure that enables semianalytic computations of large scale basket portfolio products such as CDOs. The present paper focussed entirely on the mathematical properties of this modeling approach, and leaves interesting implementation questions such as calibration and applications in portfolio credit VaR and CDO pricing as subjects for future work. 

\bigskip
\noindent{\bf Acknowledgments:\ } I am grateful to Alexey Kuznetsov and Zhuowei Zhuo for helpful discussions on the general theory of time changed Brownian motions.

\bibliographystyle{plain}

\end{document}